\documentclass[12pt,,letterpaper]{JHEP3}
\usepackage{graphics}
\usepackage{epsfig}
\usepackage{slashed}

\title{Dynamic Mott gap from holographic fermions in geometries with hyperscaling violation}

\author{ZhongYing Fan\\
Department of Physics, Beijing Normal University, 100875 Beijing,
China\\
\email{zhyingfan@gmail.com}
}

\abstract{We investigate a dynamically generated Mott gap from holographic fermions in asymptotical geometries with hyperscaling violation by employing a bulk dipole coupling for fermions. We find that when the coupling strength increases, the spectral function first appears at the negative frequency region but is soon transferred to the positive region. A stable gap and two bands emerge for all momentums when the coupling strength exceeds a critical value. Generally, the upper band on the positive frequency axis is much sharper than the lower band on the negative side. When the diploe coupling increases further, the gap becomes larger. The upper band keeps sharp while the lower band disperses and widens, concentrating on the small momentum space. We also find that the bands will be smoothed out gradually with the increasing of hyperscaling violation.  }

\preprint{}
\begin{document}

\section{Introduction}

In recent years, AdS/CFT correspondence has been widely used to study condensed-matter theory (AdS/CMT). The strongly coupled conformal theory in the boundary is mapped to weakly coupled gravity theory in the bulk. With this great advantage, people have successfully constructed holographic models of
Fermi and non-Fermi liquids in kinds of geometries\cite{1,2,3,4,5,6,7,8,9} and analytically investigated the liquids properties, showing the dispersion relation and the width of the quasiparticle-like excitation.

Since condensed-matter systems are usually described by non-relativistic field theories, in order to search more proper gravity duals people have further generalized the correspondence to non-relativistic holography, which describes anisotropic scaling behaviors for temporal and spatial coordinates\cite{10,11,12,13} i.e. Lifshitz-like geometry with dynamical exponent. For realistic systems, another important exponent i.e. hyperscaling violation will emerge and play a crucial role in low energy physics. It is certainly necessary to extend holography to this non-trivial case. This is realized by employing the standard Einstein-Maxwell-dilaton action in the bulk\cite{4,14,15,16,17}. The metric behaves like:

\begin{equation} ds^2=-\frac{dt^2}{r^{2m}}+r^{2n}dr^2+\frac{dx_i^2}{r^2} \label{1}\ , \end{equation}

where $i=1,2,...,d$ is the space index, $m$ and $n$ are related to dynamical exponent $z$ and hyperscaling violation $\theta$ by

\begin{equation} z =\frac{m+n+1}{n+2}\ ,\quad   \theta =\frac{n+1}{n+2}\cdot d\ .        \label{2}\end{equation}

Note that when $n=-1$, the metric reduces to the pure Lifshitz spacetime and $n=-2$ corresponds to a class of spacetime conformally related to $AdS_2\times R_d$ with locally critical limit $z\rightarrow \infty,\   \theta\rightarrow -\infty, \mbox{while}\  z/\theta$ is fixed to be a constant\cite{18}. The metric transforms as

\begin{equation}
t \rightarrow \lambda^z t,\ x_i\rightarrow \lambda x_i,\ r\rightarrow \lambda^{(d-\theta)/d} r,\ ds\rightarrow \lambda ^{\theta/d} ds\ .\label{3}\end{equation}

Clearly, the metric is not scale invariant. In general, the dual boundary theory exhibits this peculiar behavior below some non-trivial dimensional scale. However, we will not consider this complication. Instead we assume that the metric is asymptotical geometry with hyperscaling violation in this paper.

We found remarkable influence of hyscaling violation on the dynamical gap by introducing a magnetic dipole coupling for bulk fermions. It was first proposed in \cite{19,20} and further studied in \cite{21,22,23,shangyu}. A gap in the spectral function was opened when the dipole coupling strength $p$ exceeds some critical value, which behaves like a Mott insulator. The gap becomes wider when $p$ increases. The coupling strength $p$ plays a role similar to the dimensionless interaction strength $U/t$ in the Hubbard model of fermions. The novel feature we find is that two bands exist in the spectral function, an upper band on the positive frequency axis and a lower band on the negative side, respectively. These two bands behave qualitatively different with the increasing of the interaction $p$ and hyperscaling violation $\theta$.

The remainder of this paper is organized as follows: In section 2, we briefly review the effective gravity model, i.e, Einstein-Maxwell-dilaton theory for geometries with hyperscaling violation. In section 3, we study the bulk fermions with a dipole interaction, deriving the equations of motion for the retarded correlator.
In section 4, we numerically solve the equations of motion under proper boundary conditions and extract the main results of the emergence of the gap.
Finally, we present a conclusion in section 5.

\section{Effective Gravity Model}

The standard Einstein-Maxwell-dilaton (EMD) action reads

\begin{equation} S=\int \mathrm{d}^{d+2}x \sqrt{-g}\ [R-2(\partial{\phi})^2-V(\phi)-\frac{\kappa^2}{2}Z(\phi)F^2-\frac{\kappa^2}{2}H^2]\ ,
\label{action}\end{equation}
where the AdS radius has been set to 1. The solutions with hyperscaling violation are listed in the following:

\begin{equation} ds^2=-r^{-2m}h(r)dt^2+r^{2n}h^{-1}(r)dr^2+\frac{dx^2_i}{r^2}\ ,\qquad h(r)=1-(\frac{r}{r_h})^{\delta}\ , \label{ds2}\end{equation}

\begin{equation} F^{rt}=F_0r^{(m-n+d)}Z^{-1}(\phi)\ ,\qquad H^{rt}=H_0r^{(m-n+d)}\ , \label{fh}\end{equation}

\begin{equation} \phi=k_0\log{r}\ ,\qquad \quad k_0=\sqrt{\frac d2 (m-n-2)}\ , \label{phi}\end{equation}

\begin{equation} V(\phi)=-V_0 e^{-\beta \phi}\ ,\qquad V_0=\delta(m+d-1)\ ,\qquad  \beta=\frac{2(n+1)}{k_0}\ , \end{equation}

\begin{equation} Z^{-1}(\phi)=Z_0 e^{-\alpha \phi}+Z_1\ ,\ \alpha=\frac{2(n+d+1)}{k_0}\ ,\ Z_0=\frac{\delta(m-1)}{\kappa^2F_0^2}\ ,\ Z_1=-\frac{H_0^2}{F_0^2}\ . \end{equation}
where $\delta=m+n+d+1$, $r_h$ is the location of the horizon, $F_0$, $H_0$ are constants which are proportional to the conserved charges carried by the black brane. The Hawking temperature and the entropy density of the black brane are give by

\begin{equation} T=\frac{\delta}{4\pi } \frac{1}{r_h^{(m+n+1)}}\quad ,\quad s=\frac{1}{8\kappa^2 r_h^d}\ . \label{t}\end{equation}

In the zero temperature limit ($r_h\rightarrow \infty$), the entropy density approaches to zero. This is important for realistic systems with degenerate ground states.

In order to admit a stable theory, the dilaton solution is required to be real, leading to $m \geq n+2$ or equivalently $z \geq 1+\theta/d,\ \theta < d$. Moreover, in the asymptotic limit $r\rightarrow 0$, the field strength $F^{\mu\nu}$ diverges such that the dual chemical potential cannot be well defined. Therefore we introduce another gauge field $H=dB$ to obtain a proper definition for the finite density

\begin{equation} B(r)=\mu (1-\frac{r^{(d-m+n+1)}}{r_h^{(d-m+n+1)}})\ dt\ , \label{b} \end{equation}
where $\mu$ is the chemical potential. The constraint condition which makes $B \ \mbox{and}\ H $ finite in the UV limit is

\begin{equation} 2\leq m-n \leq d\ ,\qquad d\geq 3\ . \label{mn}\end{equation}

The divergent behavior of the field $F^{\mu\nu}$ certainly needs to be treated properly in a holographic renormalization procedure which we will not discuss in this paper. Since the bulk fermions we consider don't couple to the dilaton and $F$ fields directly, the results we obtain are still credible, in the absence of a full treatment of the holographic EMD theory.

\section{Holographic fermion with magnetic dipole coupling}

In order to explore the effects of magnetic dipole coupling on the spectral function of fermions, we start from the following action

\begin{equation} S_f[\Psi]=i \int \mathrm{d}^{d+2}x\sqrt{-g}\ \overline{\Psi}(\Gamma^a\mathcal{D}_a-M-i p \mathcal{H})\Psi+S_{bdy}[\Psi]\ , \label{faction}\end{equation}

\begin{equation} S_{bdy}[\Psi]=i\int_\epsilon \mathrm{d}^{d+1}x\sqrt{-g_\epsilon}\sqrt{g^{rr}}\overline{\Psi}_+\Psi_-\ ,\label{bdy}\end{equation}
where $S_{bdy}$ is a boundary action to ensure a well defined variational principle\cite{24} for the total fermion action. $\overline{\Psi}=\Psi \Gamma^t$, $\mathcal{D}_a=(e_a)^\mu D_\mu$, with $D_\mu=\partial_\mu-i q B_\mu+\frac 14 \omega_{\mu ab}\Gamma^{ab}$, $\Gamma^{ab}=\frac 12 [\Gamma^a, \Gamma^b]$. $\omega_{\mu ab}$ is the spin connection 1-form and $\mathcal{H}=\frac 12 \Gamma^{ab} (e_a)^\mu(e_b)^\nu H_{\mu\nu}$. $\Gamma^a$ are the $d+2$ dimensional gamma matrices; $(e_a)^\mu$ are vielbeins and $M$ is the fermion mass. Furthermore, $g_\epsilon$ is the determinant of the induced metric on the constant $r$ slice, $r=\epsilon$. $\Psi_\pm$ is defined by

\begin{equation} \Psi_\pm=\frac 12 (1\pm \Gamma^r) \Psi\ ,\qquad \Gamma^r \Psi_\pm=\pm \Psi_\pm\ .  \end{equation}

The Dirac equation derived from the action reads

\begin{equation} (\Gamma^a\mathcal{D}_a-M-i p \mathcal{H})\Psi=0 \label{dirac}\ . \end{equation}

Taking a Fourier transformation

\begin{equation} \Psi(r,x_\mu)=(-gg^{rr})^{-\frac 14}e^{-i\omega t+ik_ix^i} \psi(r,k_\mu)\ ,\quad k_\mu=(-\omega,\ \vec{k})\ ,  \end{equation}

where the prefactor was introduced to remove the spin connection in the equations of motion. Since the theory is rotational invariant, we can choose the momentum along $x_1$ direction. The Gamma matrices are chosen as follows

\begin{equation} \Gamma^r=
\left( \begin{array} {ccc}
   -\sigma^3 & 0\\
   0 & -\sigma^3
\end{array}\right)
\ ,\quad
\Gamma^t=
\left( \begin{array}{ccc}
  i\sigma^1 & 0 \\
  0 & i\sigma^1
\end{array}\right)
\ ,\quad
\Gamma^{x_1}=
\left( \begin{array}{ccc}
    -\sigma^2 & 0 \\
    0 & -\sigma^2
\end{array}\right)\ ,
\end{equation}
where $\sigma$ are Pauli matrices. We further set

\begin{equation}
\psi= \left(\begin{array}{ccc} \psi_+ \\ \psi_-\end{array}\right)\ ,\qquad \psi_\pm=\left(\begin{array}{ccc}u_\pm \\ d_\pm \end{array}\right)\ .
\end{equation}

Since the Dirac equation is first order, there exists some relation between $\psi_+$ and $\psi_-$. Assuming $\psi_+(r,k_\mu)=- i \xi(r,k_\mu)\psi_-(r,k_\mu)$, we can derive an elegant equation to extract correlators

\begin{equation}\label{xi}
\sqrt{g^{rr}}\partial_r \xi_\pm +2M \xi_\pm=(v_-  \pm k\sqrt{g^{x_1x_1}})\xi^2_\pm+(v_+ \mp k\sqrt{g^{x_1x_1}})\ ,
\end{equation}
where $\xi_+=iu_-/u_+,\ \xi_-=id_-/d_+$, $\xi_\pm$ are the eigenvalues of the matrices $\xi$. $v_\pm$ are defined as follows:

\begin{equation} v_{\pm}=\sqrt{-g^{tt}}(\omega+q B_t \pm p \sqrt{g^{rr}}\partial_r B_t)\ . \end{equation}

The corresponding retarded functions can be readily obtained as follows\cite{25}

\begin{equation} G_O(k_\mu)=\lim_{r\rightarrow 0} \xi(r,k_\mu)\ . \label{twopointfunction}\end{equation}

At the event horizon, we impose in-falling boundary conditions

\begin{equation} \xi(r_h,k_\mu)=i\ ,\qquad \mbox{for}\ \omega\neq 0\ . \label{eventcondition} \end{equation}

We emphasize that the dimension of the fermionic operator $O$ is $\Delta=(m+d)/2$ which leads to the fact that the unitarity bound was automatically satisfied with $m\geq 0$ given by the null energy conditions\cite{25}. The fermion mass decouples from the operator UV dimension, contributing only to the IR physics which is peculiar in the asymptotical geometries with hyperscaling violation.
\section{Numerical results and Emergence of the gap}

To extract the effects of bulk dipole coupling on the spectral function, we need to numerically solve the flow equation (\ref{xi}) with initial conditions (\ref{eventcondition}). The spectral function is proportional to $ImG(\omega,k)$, up to normalization. Due to the relation $G_{11}(\omega,k)=G_{22}(\omega,-k)$, we will only consider $G_{22}(\omega,k)$ and omit the subscript in the following. For convenience, we set $M=0, \mu=1, q=2, z=2, d=3$. The dipole interaction strength $p$ and hyperscaling violation $\theta\ (\mbox{or}\ n)$ remain to be free.

First, we fix the hyperscaling violation, considering $n=0$ case. From the plots above in figure \ref{f1}, we find that a sharp quasi-particle like peak occurs at $k_F\approx 1.2044$ for $p=0$, indicating the existence of a Fermi surface. The dual liquid is of Fermi type with linear dispersion relation at the maximum height of the spectral function. For larger charge $q$, it has been investigated with great detail in \cite{25} that more branches of Fermi surfaces appear. Moreover, when $q$ is sufficiently large, there exists a peculiar Fermi shell-like structure, which contains many sharp and singular peaks in some narrow interval of the momentum space.

\FIGURE[ht]{\label{f1}
\includegraphics[width=7.5cm]{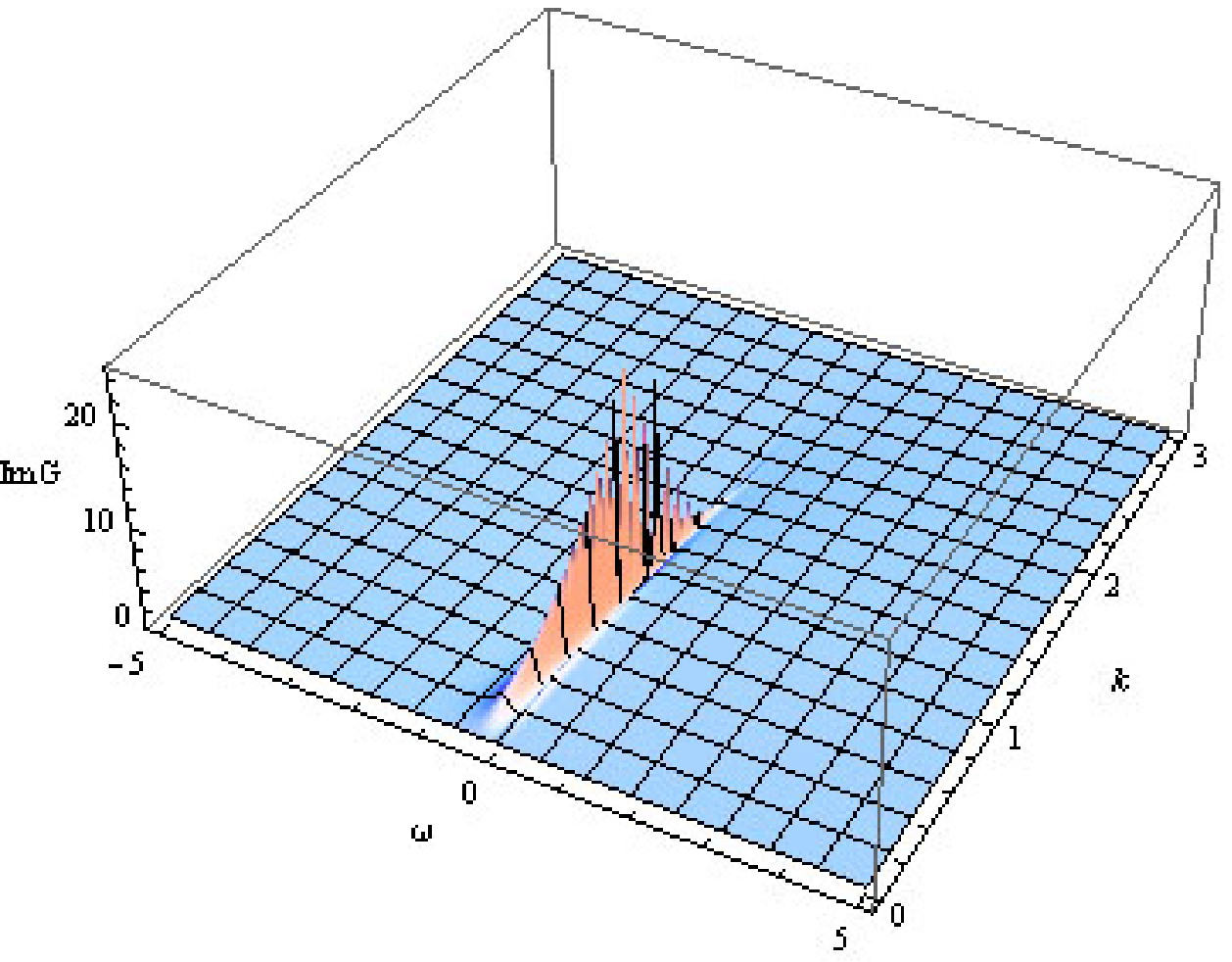}
\includegraphics[width=7.5cm]{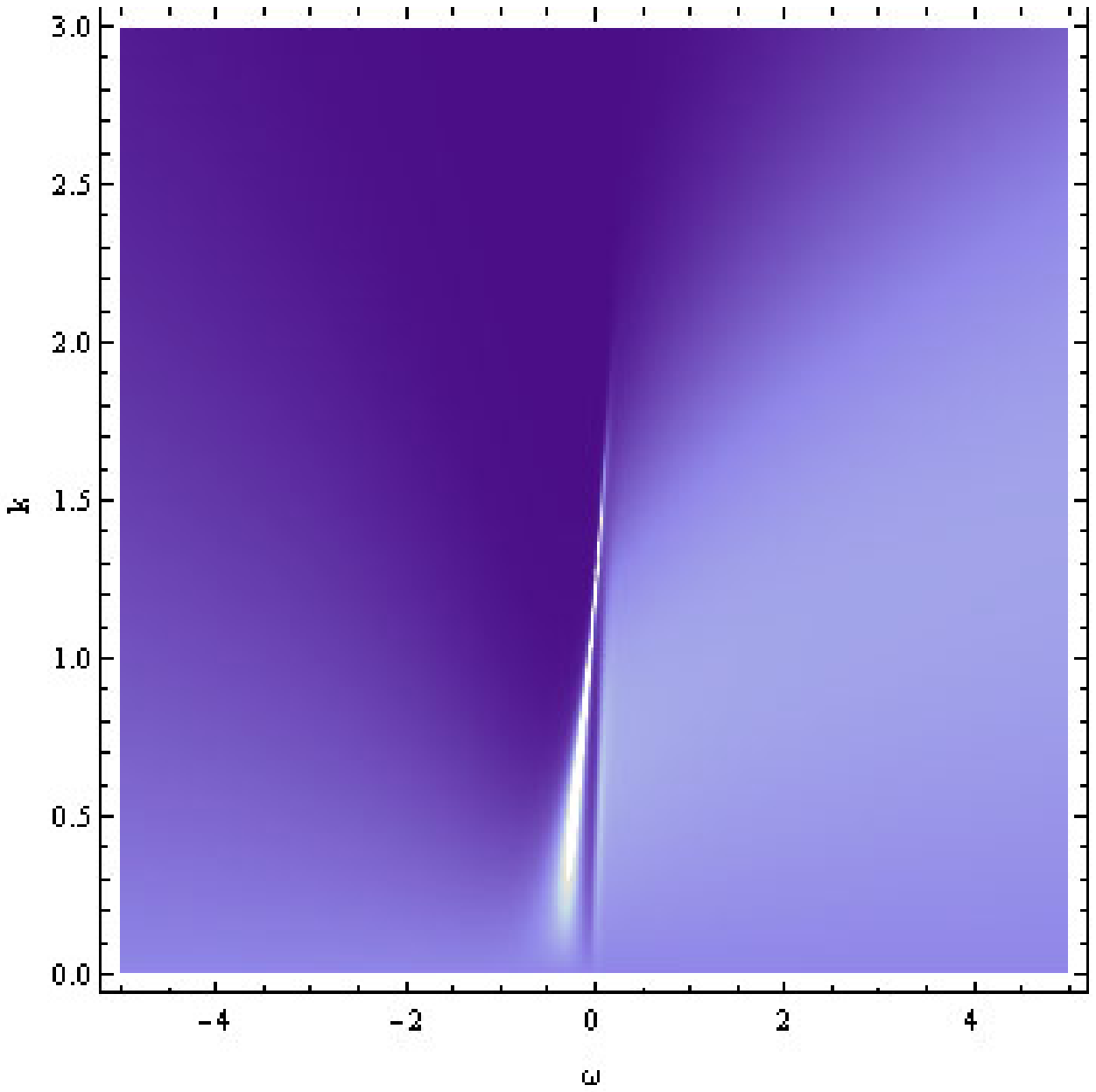}
\includegraphics[width=7.5cm]{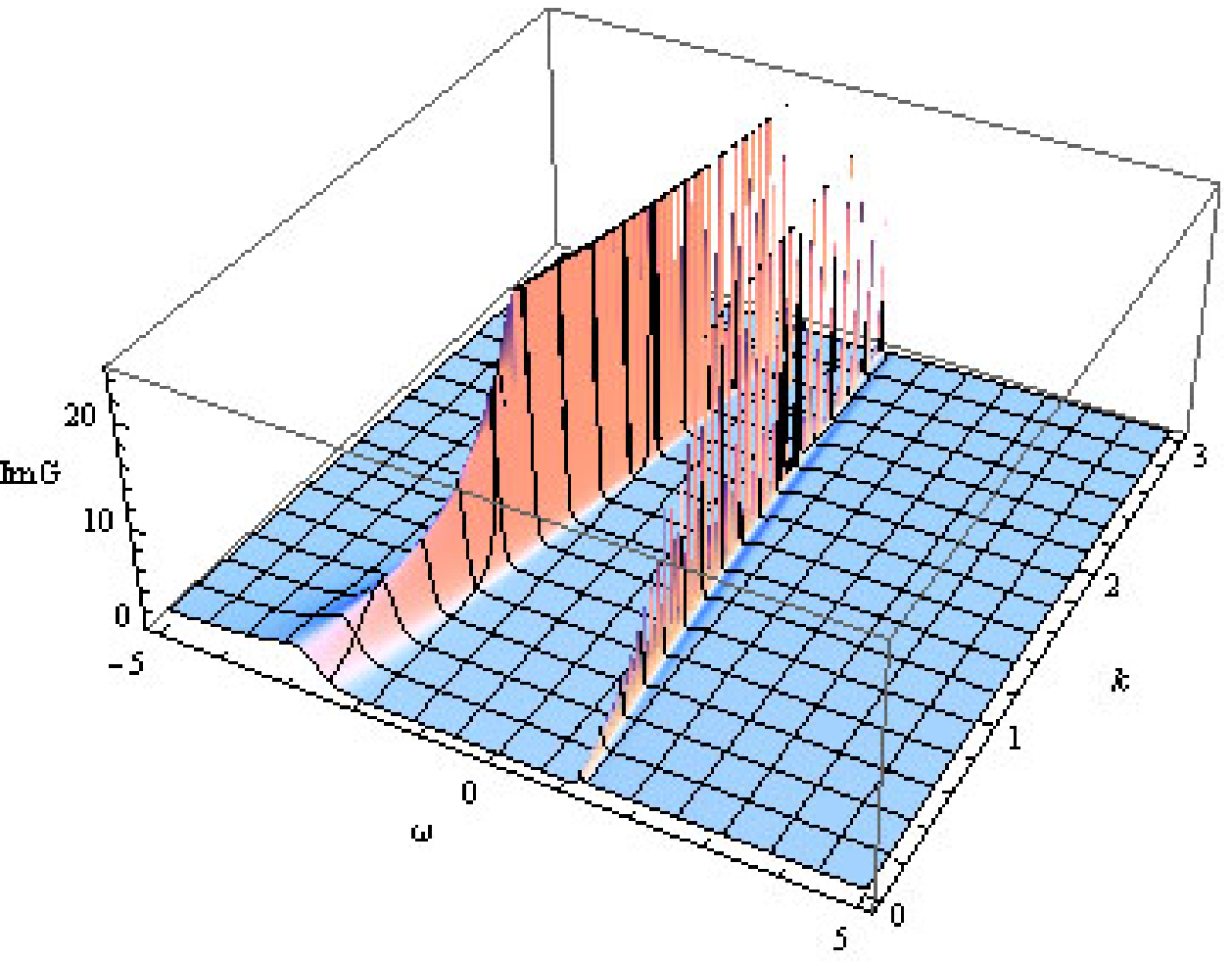}
\includegraphics[width=7.5cm]{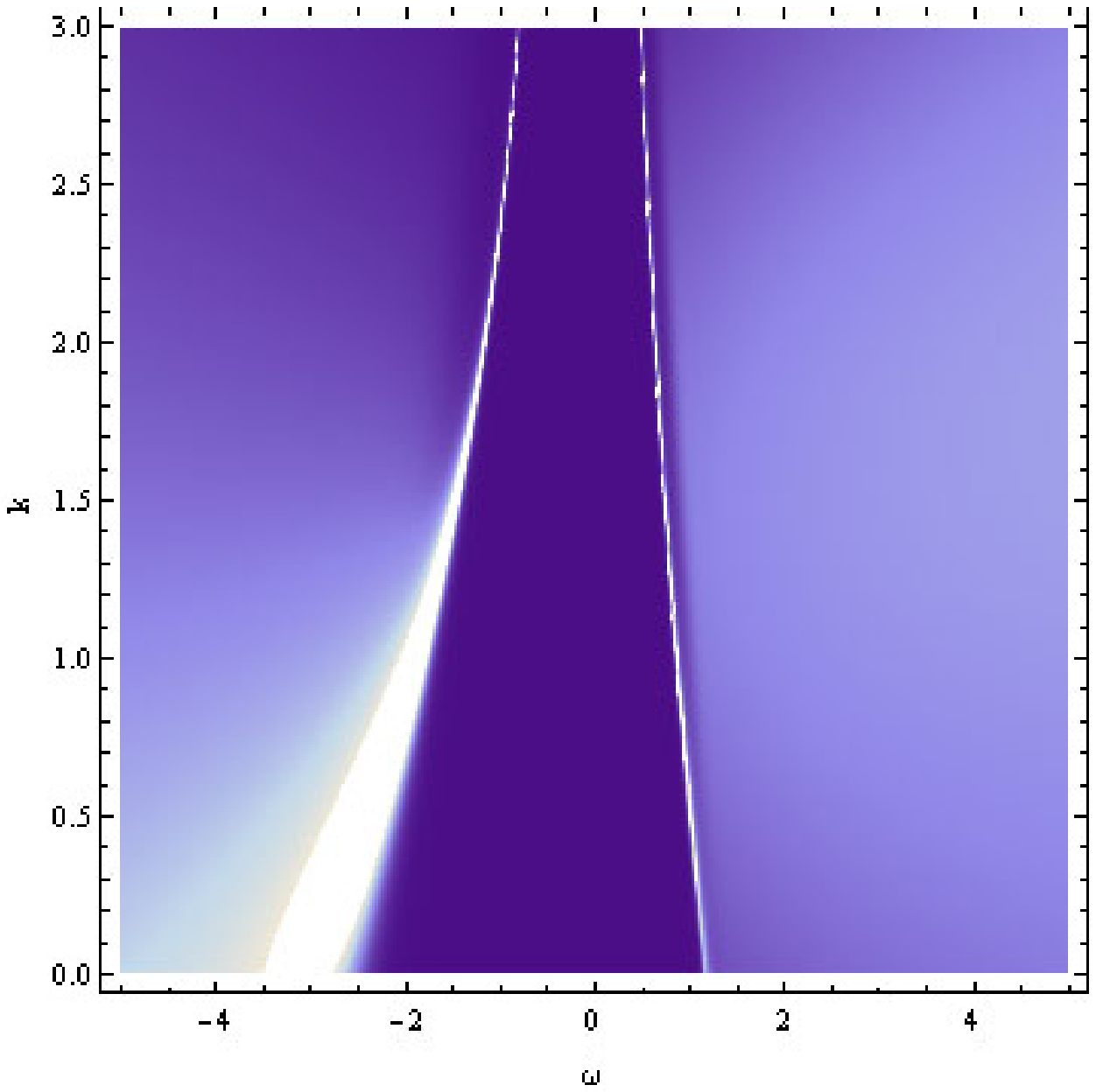}
\caption{The 3D and density plots of $ImG(\omega,k)$. In the plots above, $p=0$, a sharp peak occurs at $k\approx 1.2044$. In the plots below, $p=4$, a gap emerges around $\omega=0$.}
}

When the interaction strength was turned on at $p=4$, a gap emerges as is shown in the plots below of figure 1. There are two bands, located at positive frequency (we call it upper band)and negative frequency (called lower band) regions respectively. Evidently, the lower band is stronger than the upper one, occupying the main intensity of the spectral function. More interestingly, the upper band appears very sharp. In the big momentum region, the lower band is also as sharp as the upper one but disperses for relatively small momentum.

\FIGURE[ht]{\label{f2}
\includegraphics[width=7.5cm]{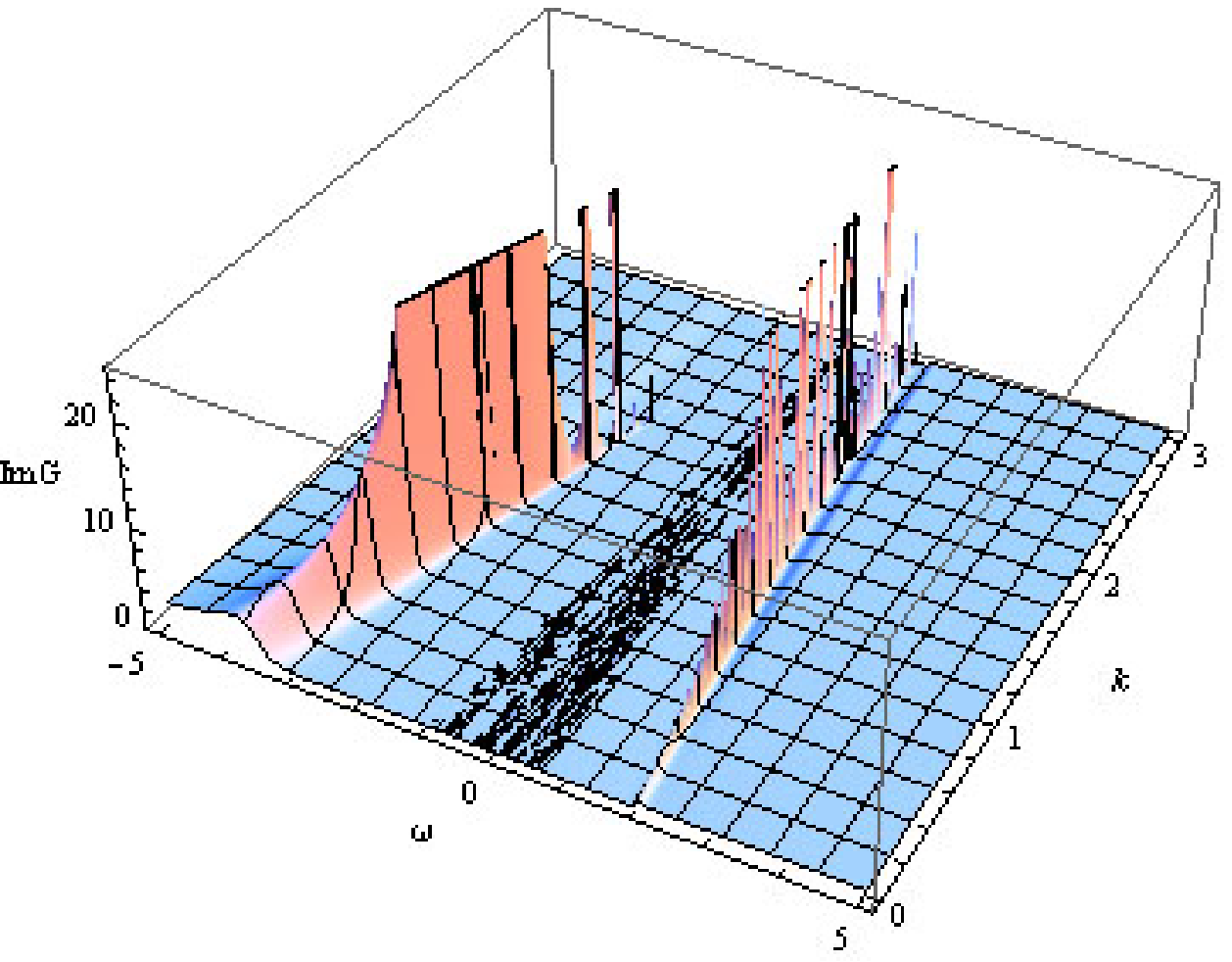}
\includegraphics[width=7.5cm]{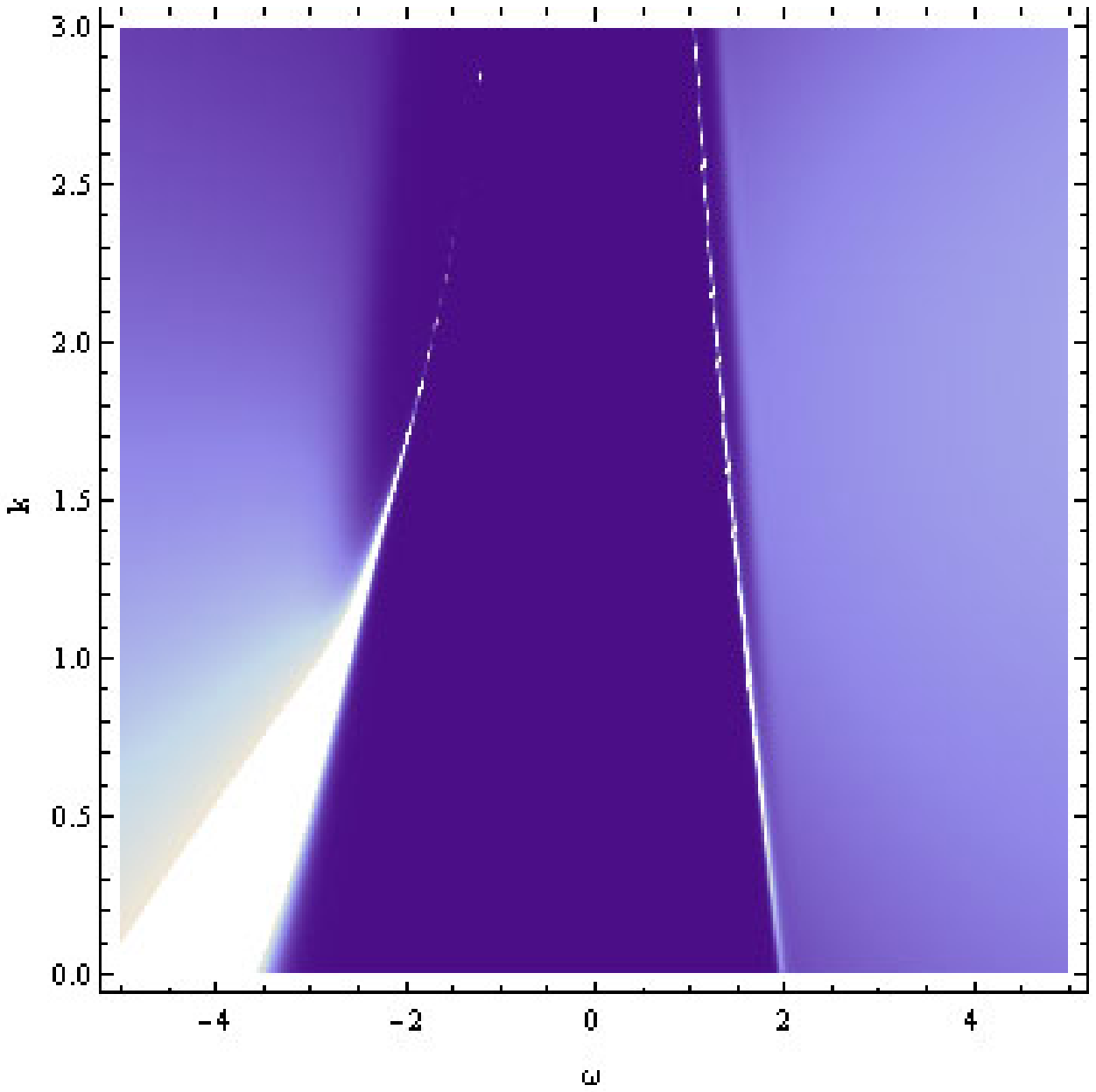}
\includegraphics[width=7.5cm]{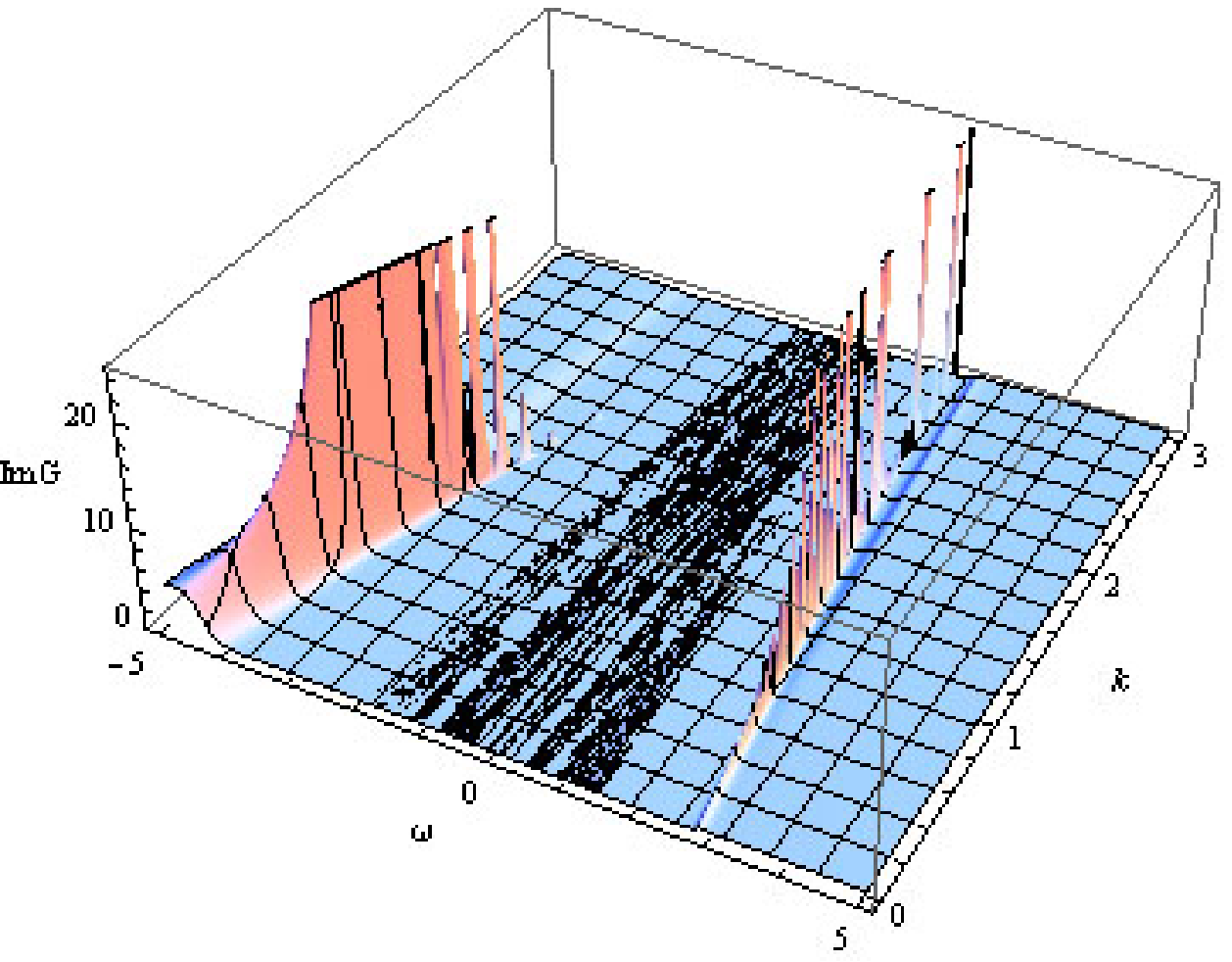}
\includegraphics[width=7.5cm]{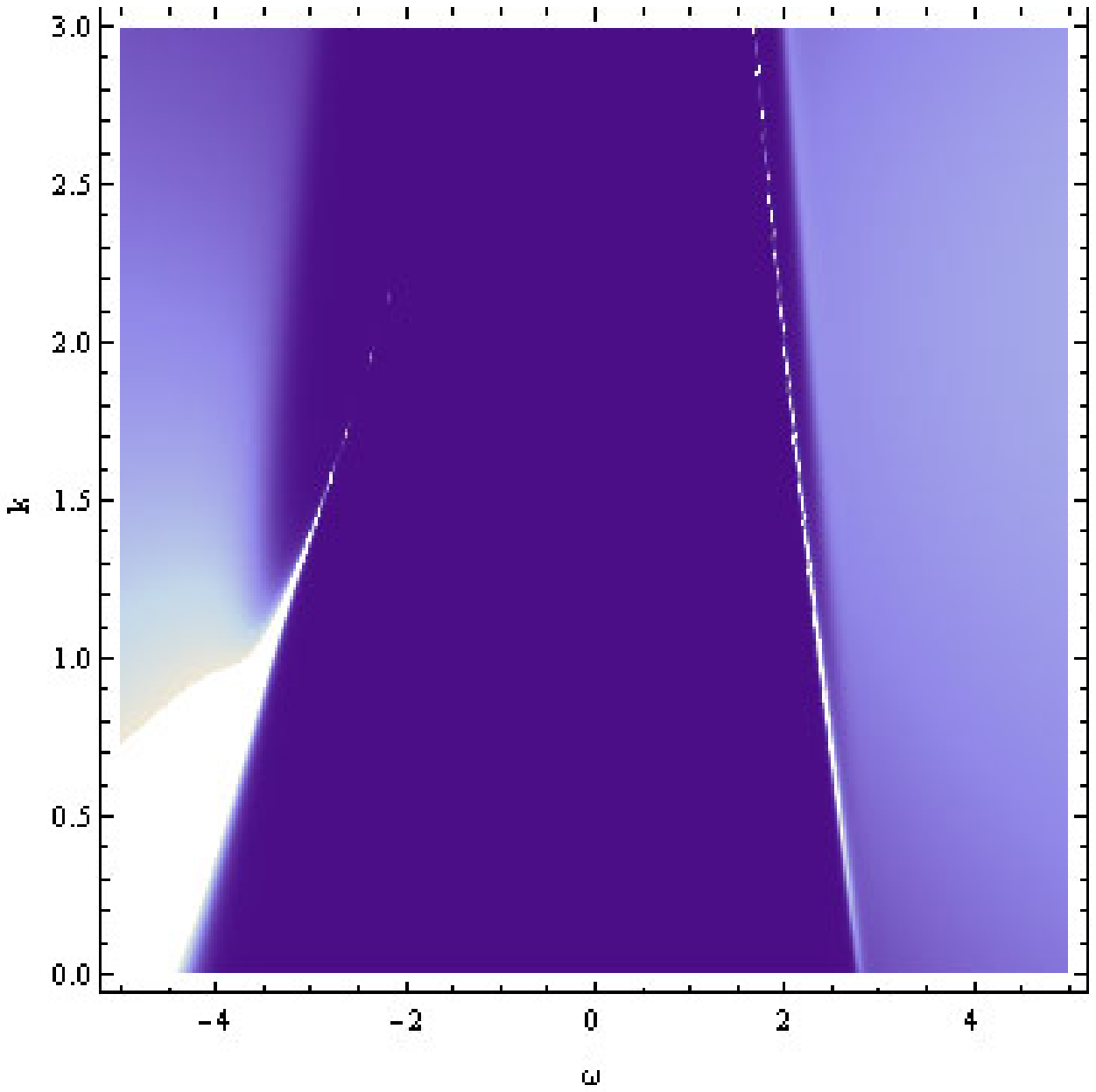}
\caption{The 3D plots and density plots of $ImG(\omega,k)$ for $p=6$ (plots above)and $p=8$ (plots below).}
}

From plots in figure \ref{f2}\footnote{The black part in the 3D plots are purely numerical noise.}, we can see that when $p$ increases further, the gap becomes larger. The upper band still keeps sharp for all momentum, translationally moving to the higher frequency region. However, the lower band is deformed much by transfer of the spectral weight to relatively higher momentum space.

\FIGURE[ht]{\label{f3}
\includegraphics[width=7.5cm]{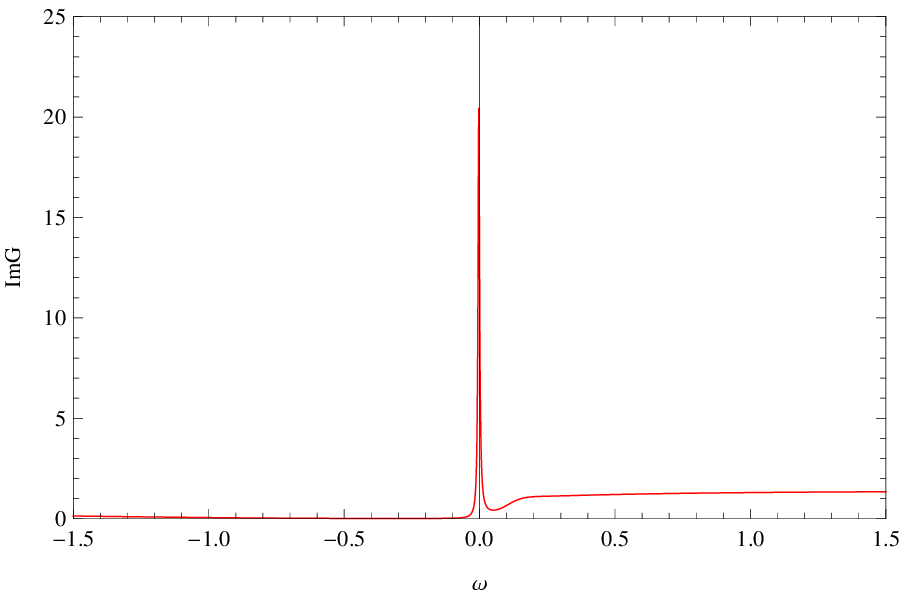}
\includegraphics[width=7.5cm]{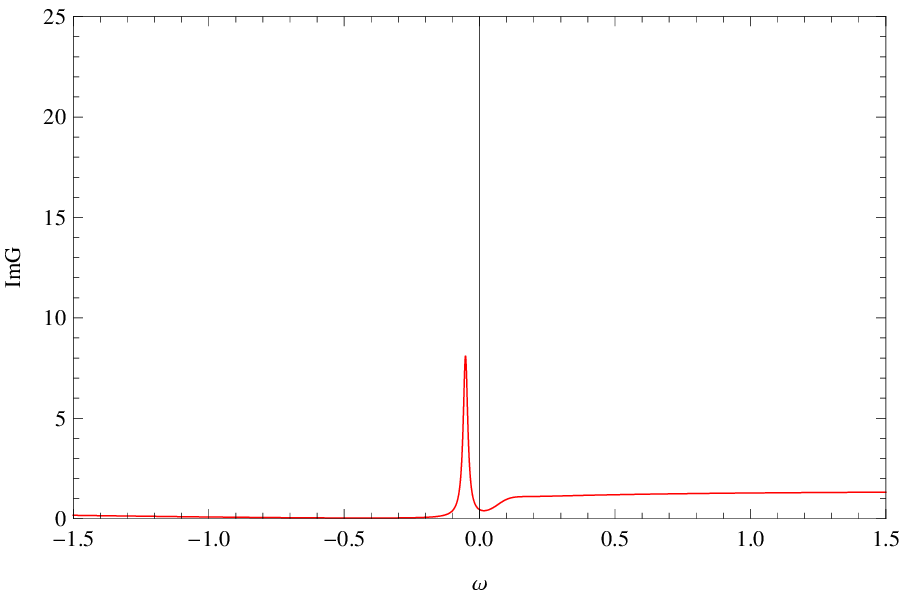}
\includegraphics[width=7.5cm]{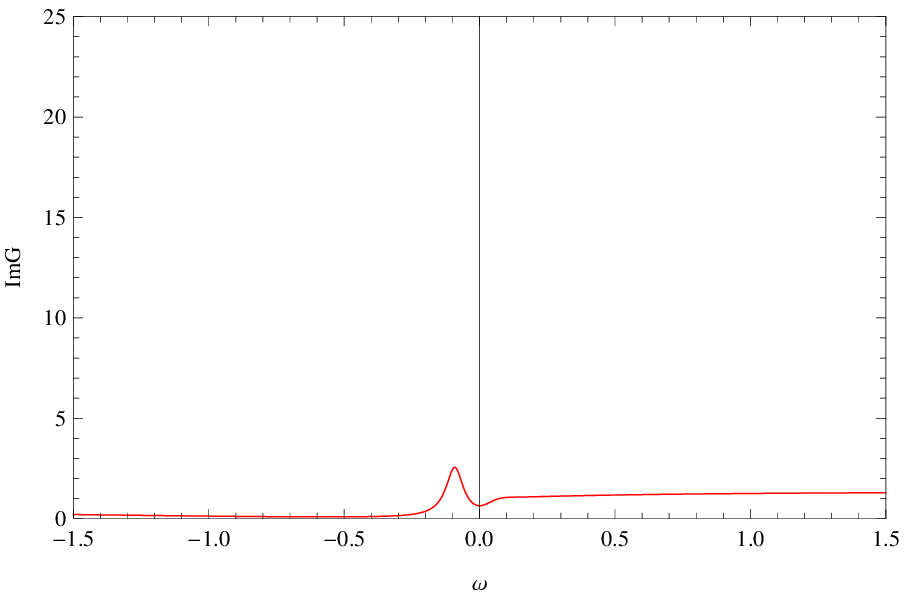}
\includegraphics[width=7.5cm]{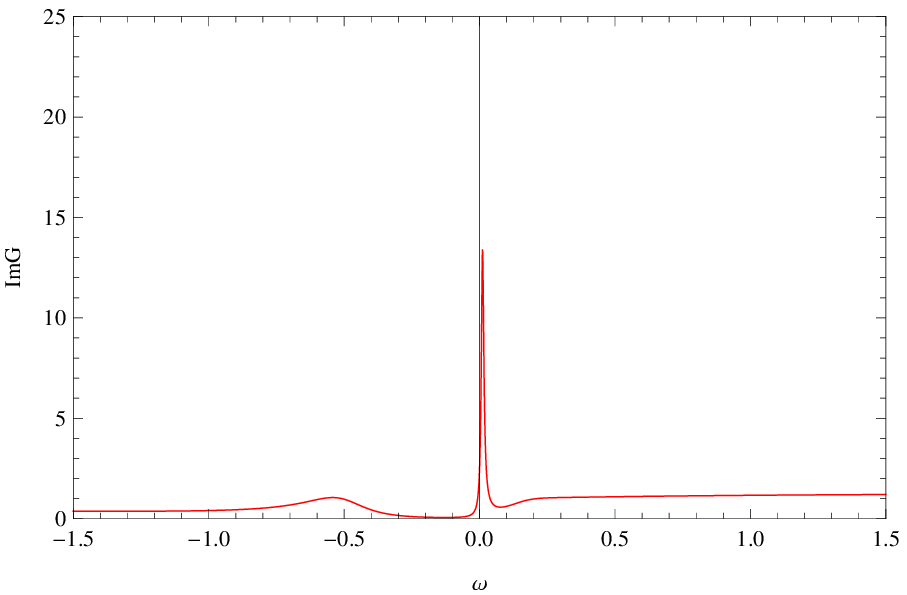}
\includegraphics[width=7.5cm]{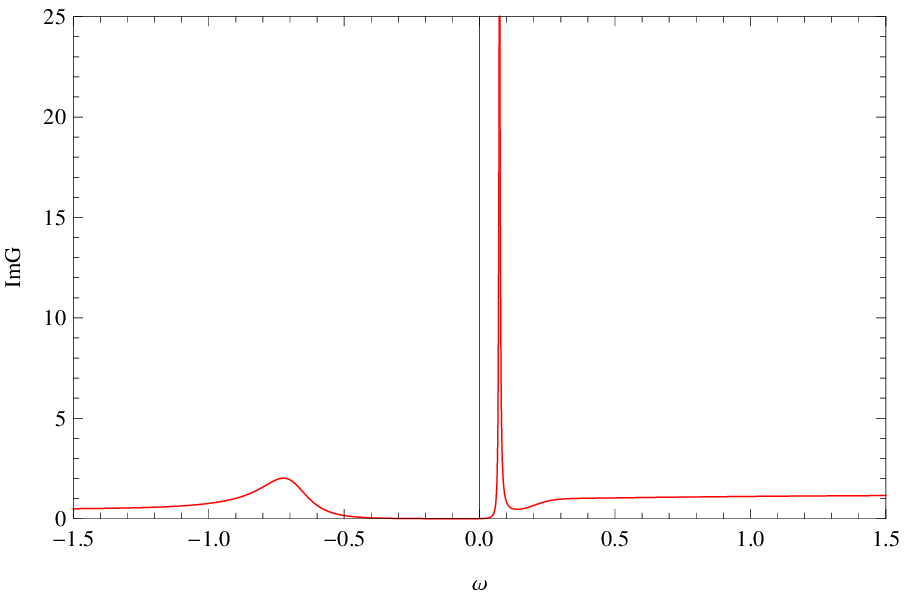}
\includegraphics[width=7.5cm]{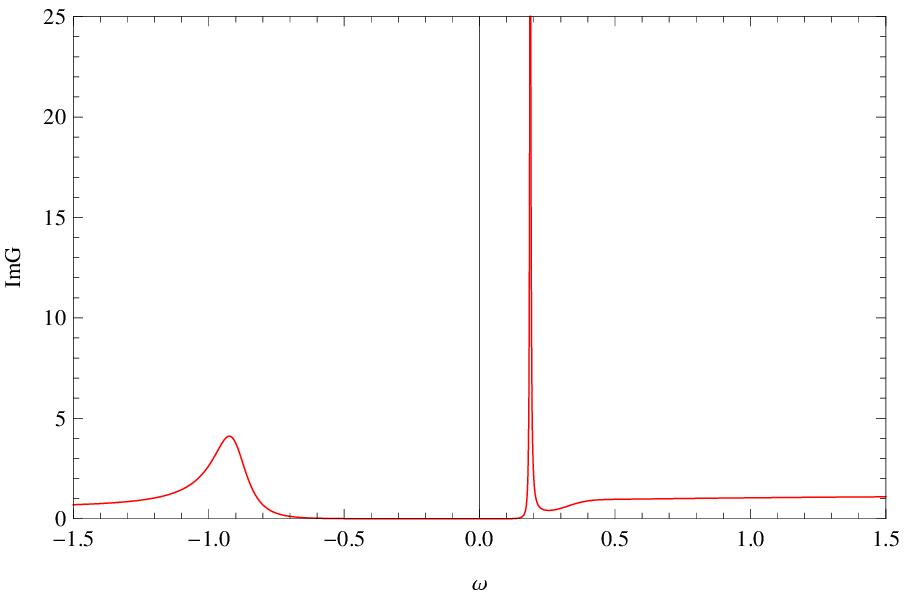}
\caption{The plots of $ImG(\omega,k)$ ($k=1.2$) for $p=0.01, 0.2, 0.4, 1.1, 1.5, 2.0$, from left to right and  top to down, respectively.}
}

In order to show the emergence of the gap in detail, we present the plots of spectral function in figure \ref{f3}.
For very small $p$, the spectral function still has a sharp peak at $\omega=0$, showing the main feature of a Fermi surface. As $p$ increases, the intensity of the peak degrades and the spectral density begins to appear at the negative frequency axis. As $p$ increases further, the spectral density is transferred to the positive frequency region. Finally, at some critical interaction strength $p_{crit}$ the original sharp peak at $\omega=0$ disappears and two stable bands emerge in both frequency regions.

\FIGURE[ht]{\label{f7}
\includegraphics[width=7.5cm]{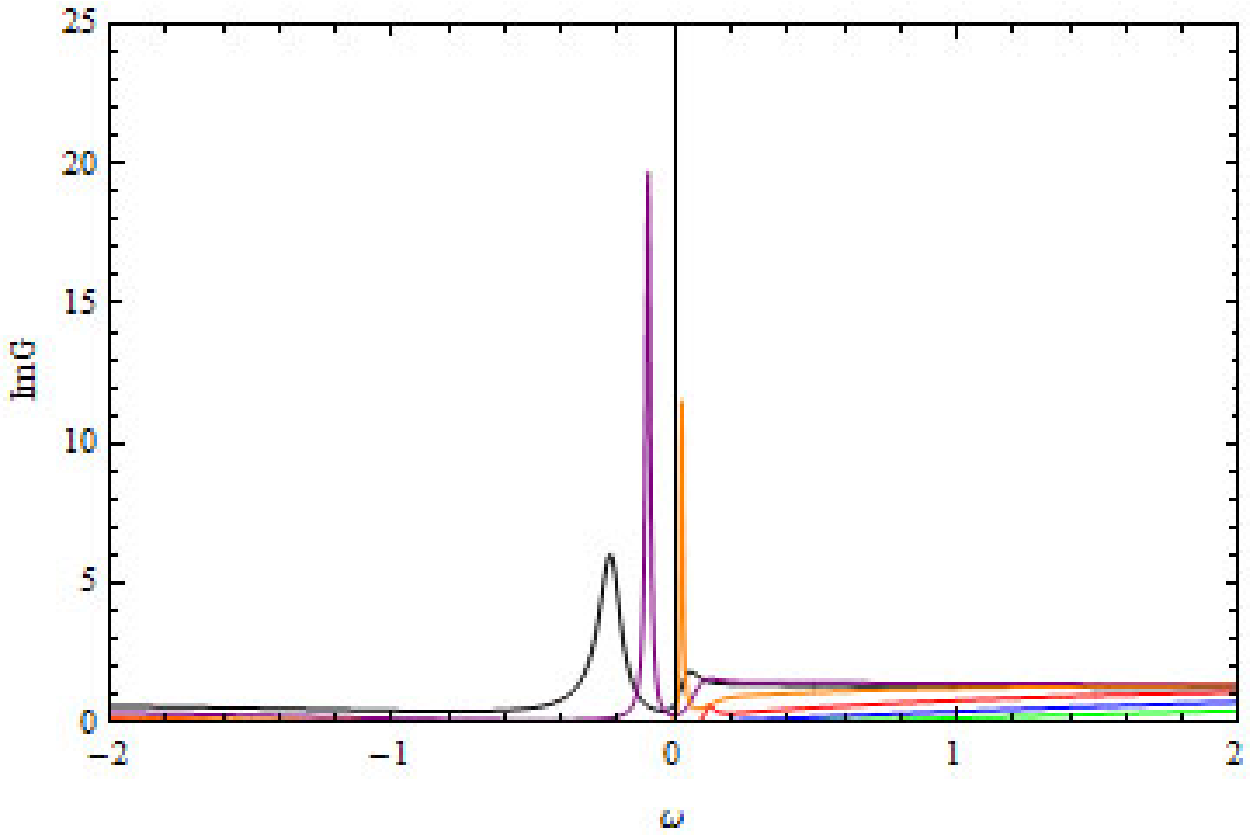}
\includegraphics[width=7.5cm]{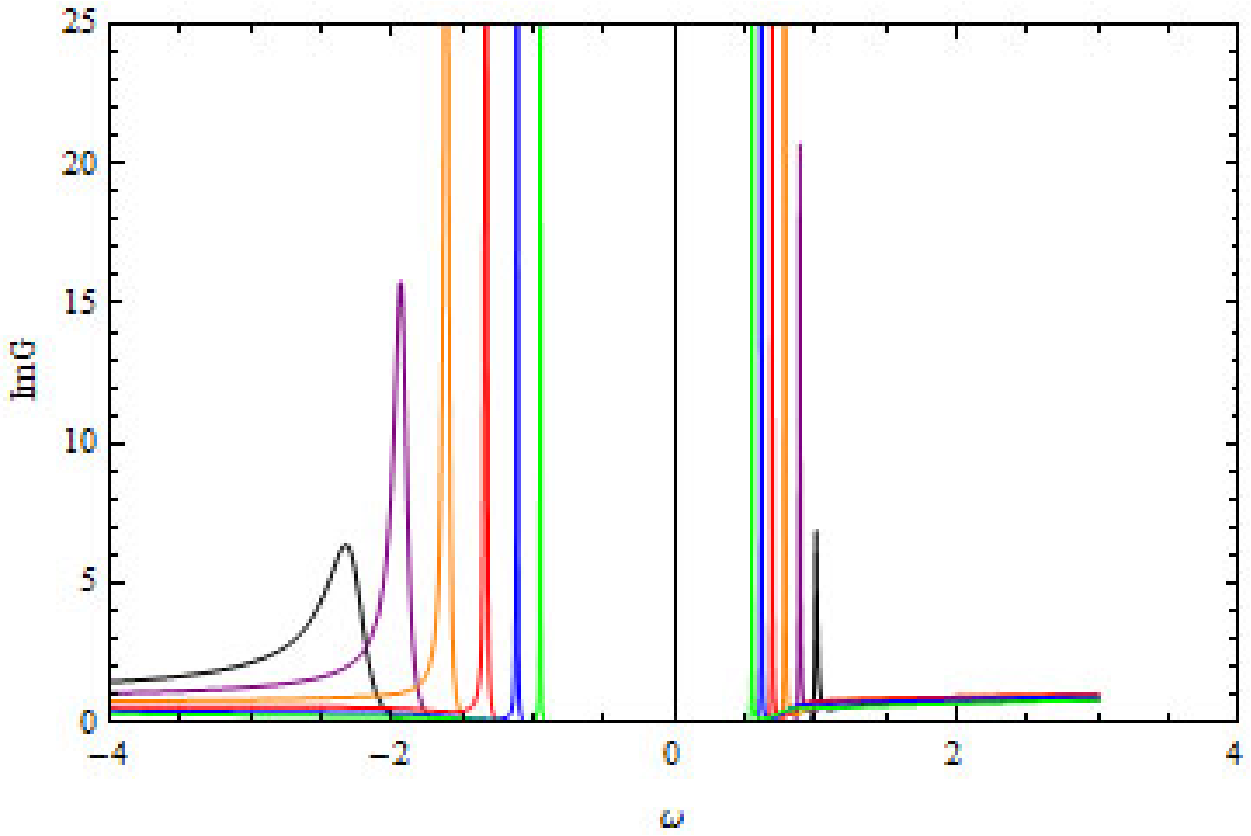}
\includegraphics[width=7.5cm]{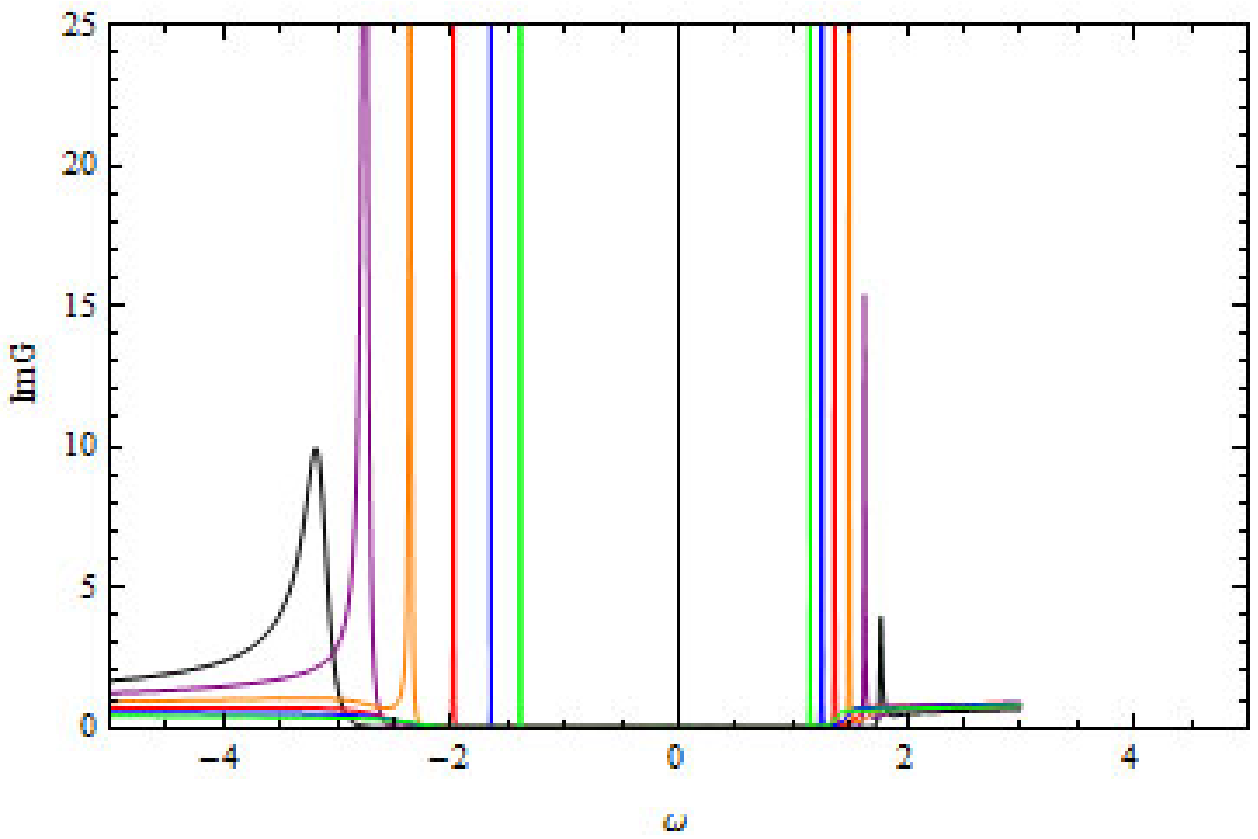}
\includegraphics[width=7.5cm]{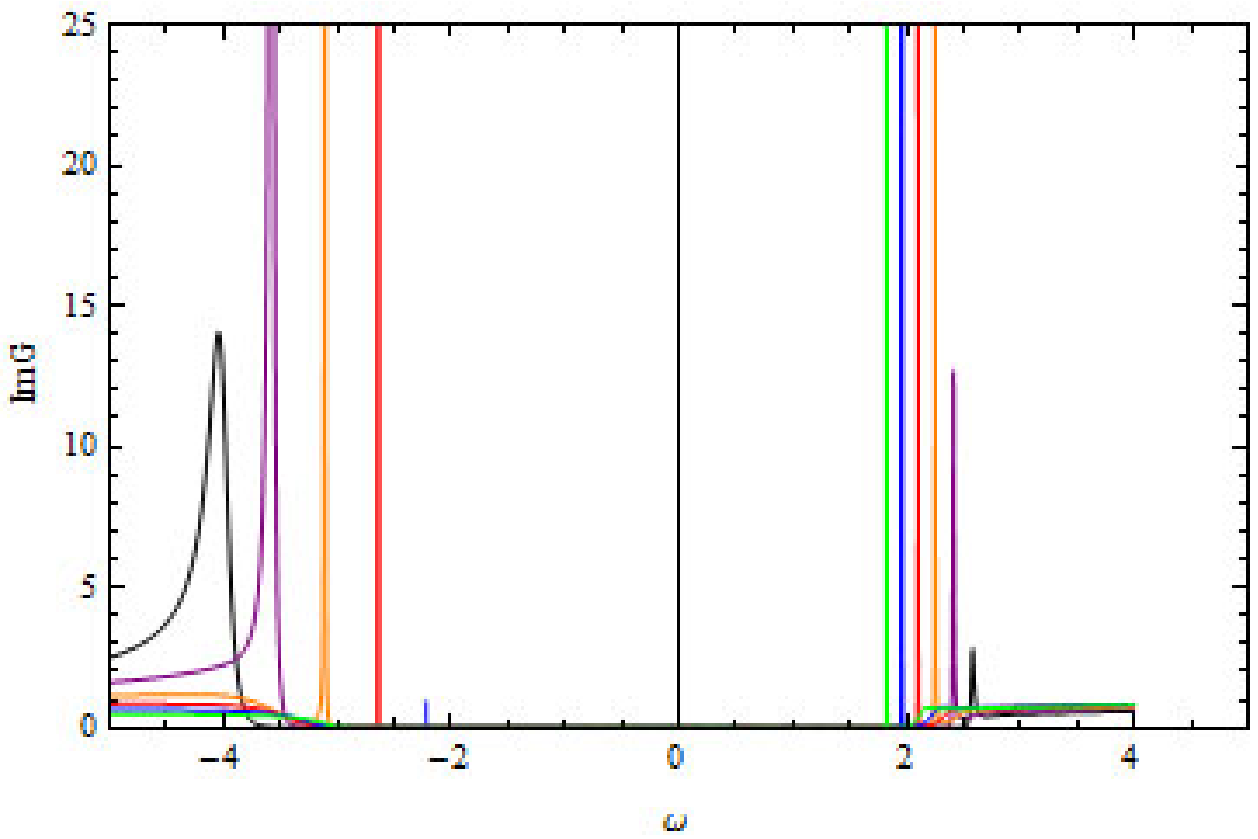}
\caption{The plots of the spectral function $ImG(\omega,k)$ as a function of $\omega$ for sample values of $k\in [0.5,2.5]$ for $p=0$ (left plot above), $p=4$ (right plot above), $p=6$ (left plot below) and $p=8$ (right plot below); $k=0.5$ (black), $k=0.9$ (purple), $k=1.3$ (orange), $k=1.7$ (red), $k=2.1$ (blue) and $k=2.5$ (green).}
}

To further explore the properties of the spectral function, we show the spectral function as a function of $\omega$ for $p=0,\ 4,\ 6,\ 8$ for sample values of momentum. From the left plot above in figure \ref{f7}, some peaks appear at both frequency regions. Around $\omega=0$, the peaks become sharper with its height tending to infinity, indicating that a Fermi surface exists at $k=k_F$. When $p$ is amplified, the quasiparticle-like peaks around $\omega=0$ degrade and vanish when $p$ exceeds some critical value $p_{crit}$. A gap will be opened for all momentum as the cases in literatures \cite{19,20,21,22}. Evidently, the upper band appears sharper than the lower one. For the lower band, the height of the spectral function increases monotonically with the increasing of momentum. As $p$ increases further, the gap widens. Both of the bands appear robust. Notice that in the right plot below of figure \ref{f7}, the lower band disappears (the green and blue lines) when momentum exceeds some critical value, implying that a redistribution and deformation happens. All of these results are consistent with our 3D and density plots in figure \ref{f2}.

In order to determine the critical strength $p_{crit}$, we plot the density of states $A(\omega)$, the total spectral weight, which is defined by the integral of the spectral function $ImG(\omega,k)$ over $k$. We find that the onset of the gap is at $p_{crit}\approx 1.2$. Notice that for small $p$ ($p< p_{crit}$) the total spectral weight mainly distributes at the negative frequency region. As $p$ increases, it transfers to the positive region to open a gap. When the value of $p$ is large enough, the spectral weight will redistribute and  backtrack to the negative region again. These results are compatible with our previous observations (figure 1, figure 2, figure 3 and figure 4).

Finally, in figure \ref{f4} we find that the width of the gap $\Delta$ increases with the increasing of the interaction strength $p$.
\FIGURE[ht]{\label{fdens}
\includegraphics[width=7.5cm]{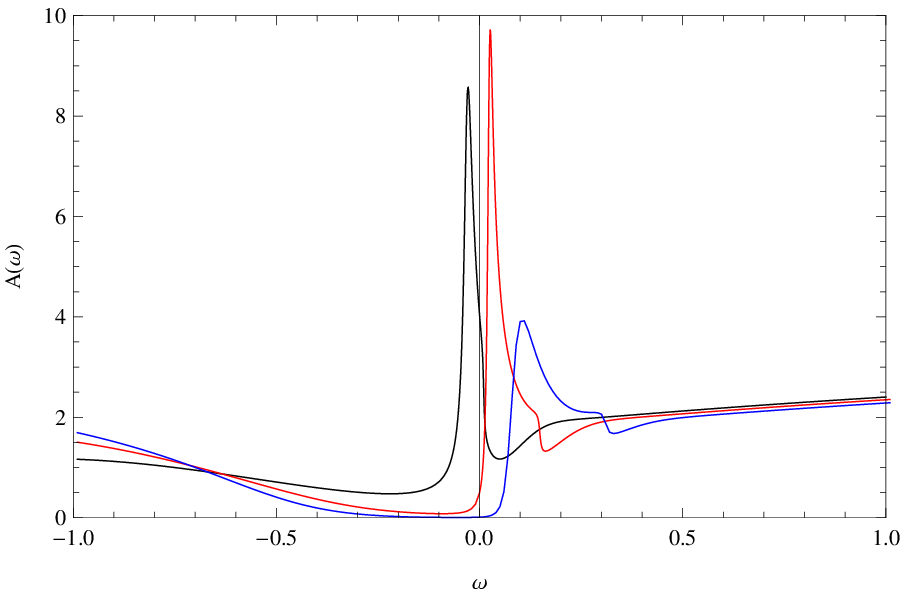}
\includegraphics[width=7.5cm]{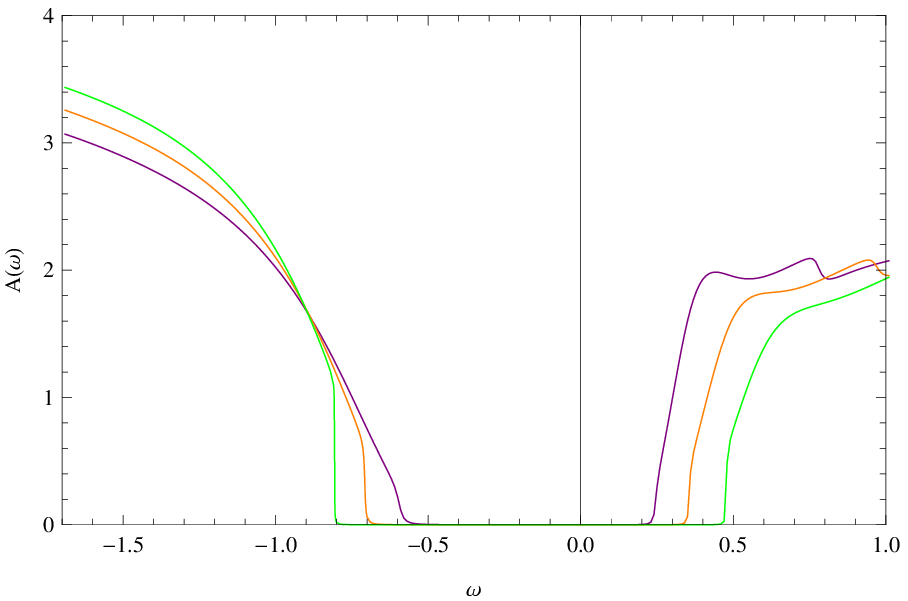}
\caption{The plots of density of states $A(\omega)$; in the left plot, $p=0.7$ (black), $p=1.2$ (red) and $p=1.7$ (blue); in the right plot, $p=3$ (purple), $p=3.5$ (orange) and $p=4$ (green); The onset of the gap is at $p\approx 1.2$. }
}

\FIGURE[ht]{\label{f4}
\includegraphics[width=9cm]{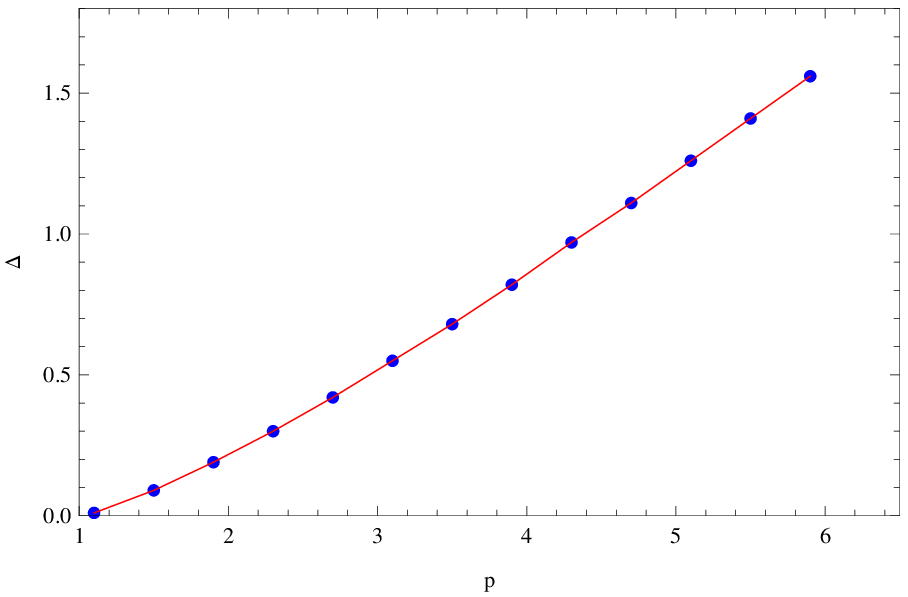}
\caption{The gap width $\Delta$ as a function of $p$.}
}

We now vary the hyperscaling violation $\theta$ with fixed $p$. Without loss of generality, we set $p=6$. In figure \ref{f5} and \ref{f6}, we show plots of the spectral function for $n=0.5$ and $n=1$. Clearly, the bands are highly suppressed as $n$ increases. The upper band disappears first. The lower band also becomes smooth gradually. As $n$ is amplified further, we may argue that in the $\theta\rightarrow d$ limit, any sharp peak of the spectral function will be completely smoothed out. The spectral density may transfer and redistribute to all frequency-momentum space homogeneously, with no explicit gap and band structure. This probably indicates some unknown critical phase.

\section{Conclusions}

\FIGURE[ht]{\label{f5}
\includegraphics[width=7.5cm]{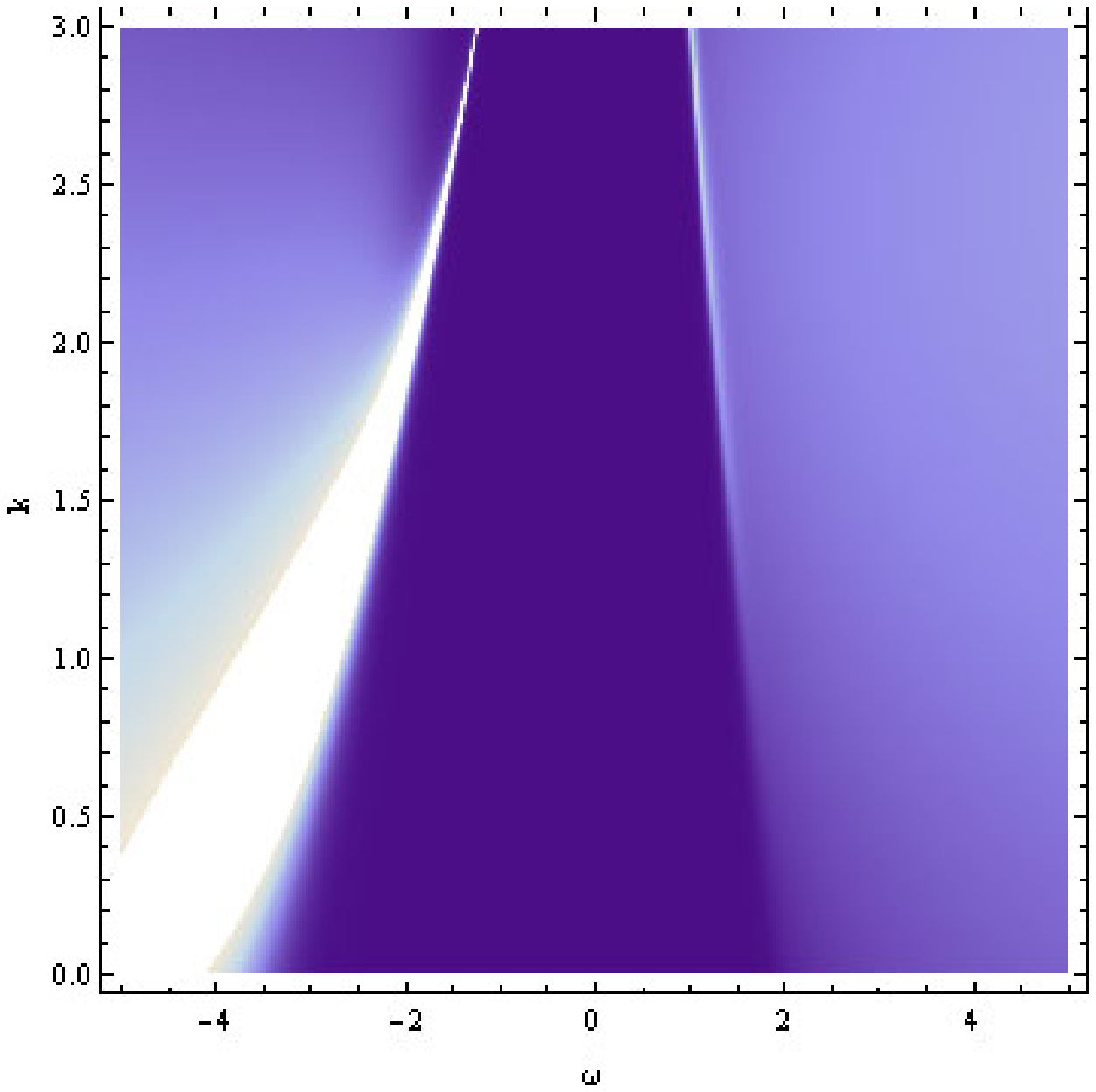}
\includegraphics[width=7.5cm]{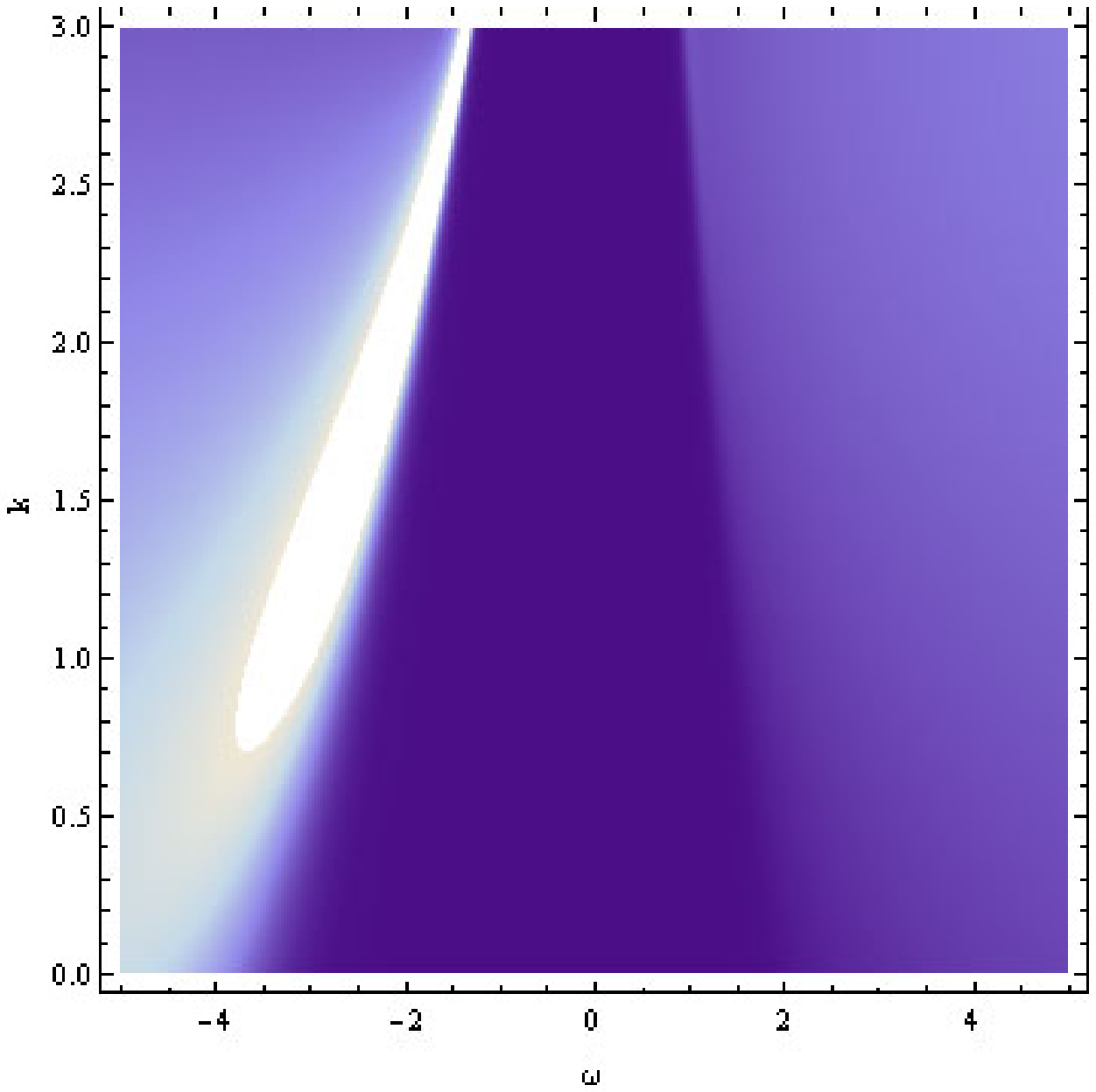}
\caption{The density plots of $ImG(\omega,k)$, $n=0.5$ for the left plot, $n=1$ for the right plot respectively. }
}

\FIGURE[ht]{\label{f6}
\includegraphics[width=7.5cm]{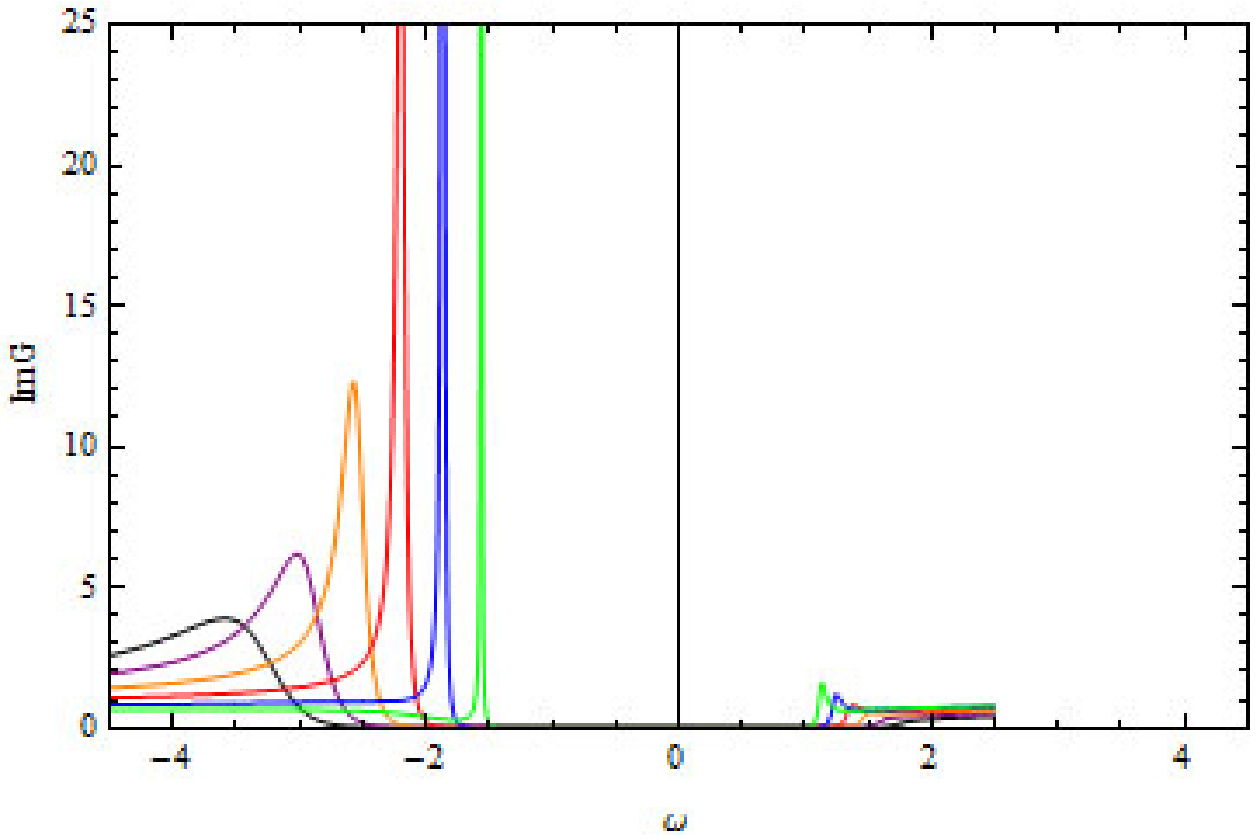}
\includegraphics[width=7.5cm]{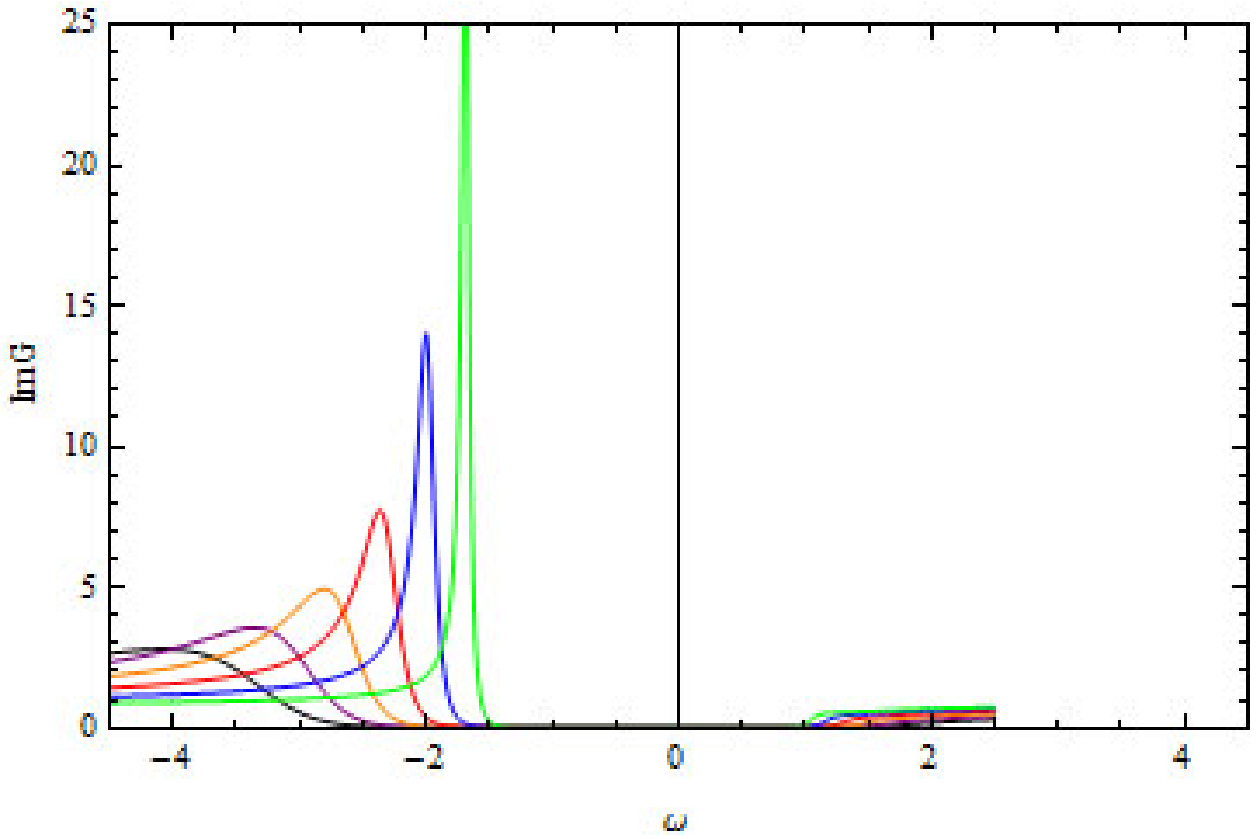}
\caption{The plots of spectral function $ImG(\omega,k)$ as a function of $\omega$ for sample values of $k\in [0.5,2.5]$ for $n=0.5$ (the left plot), $n=1$ (the right plot); $k=0.5$ (black), $k=0.9$ (purple), $k=1.3$ (orange), $k=1.7$ (red), $k=2.1$ (blue) and $k=2.5$ (green).}
}

In this paper, we have studied the novel features of fermions in the presence of bulk dipole coupling in the geometries with hyperscaling violation. For a finite hyperscaling violation $\theta=d/2$, we observe that when the dipole interaction strength $p$=0, a sharp quasi-particle like peak occurs near $k_F\approx 1.2044$ at zero frequency, showing the existence of a Fermi surface. As $p$ increases, the intensity of the sharp peak degrades and the spectral weight begins to appear at the negative frequency region but is soon transferred to the positive frequency space. When $p$ crosses a critical value $p_{crit}$, the Fermi sea disappears. Instead, a stable gap and two bands emerge for all momentums. The upper band appears sharper than the lower one which however occupies the main intensity of the spectral function. When $p$ increases further, the gap becomes wider. The upper band keeps sharp. In contrast, the lower band is deformed much by redistributing the intensity to small momentum space. We also find that the width of the gap increases with the increasing of $p$.

When we fix $p=6$ and turn on larger hyperscaling violation at $n=0.5, 1$, the peaks and bands are substantially suppressed. More interestingly, the upper band disappears first while the lower band becomes smooth gradually. Thus, the strength of the spectral density might distribute homogeneously in all frequency-momentum space in the $\theta\rightarrow d$ limit. It is of certain interests to explore this postulated critical phase in this limit. We will address it in the near future.

\section{Acknowledgments}
I would like to thank Professor Sije Gao for his useful suggestions and encouragement. This work is supported by NSFC Grants NO.10975016, NO.11235003 and NCET-12-0054.

\end{document}